\documentstyle[aps,epsf]{revtex}

\begin{document}
\title{Quantum spin tunneling in a soluble quasi-spin model: An angle-based
potential description}
\author{D. Galetti, B. M. Pimentel}
\address{Instituto de F\'{\i}sica Te\'{o}rica\\
Universidade Estadual Paulista - UNESP\\
Rua Pamplona 145\\
01405-900 - S. Paulo - S.P. - Brazil}
\author{C. L. Lima}
\address{Instituto de F\'{\i}sica\\
Universidade de S\~{a}o Paulo - USP \\
C.P. 66318 \\
05315-970 - S. Paulo - S.P. - Brazil}
\date{\today }
\maketitle

\begin{abstract}
We propose an approach which allows to construct and use a potential
function written in terms of an angle variable to describe interacting spin
systems. We show how this can be implemented in the Lipkin-Meshkov-Glick,
here considered a paradigmatic spin model. It is shown how some features of
the energy gap can be interpreted in terms of a spin tunneling. A discrete
Wigner function is constructed for a symmetric combination of two states of
the model and its time evolution is obtained. The physical information
extracted from that function reinforces our description of phase
oscillations in a potential.
\end{abstract}

 \maketitle

\section{Introduction}

Quantum tunneling is a very peculiar quantum phenomenon, and has long been
studied in a variety of contexts. Due to different aspects of the process,
coherent or incoherent tunneling have also been distinguished, according to
a possible periodicity of the motion of the particle trapped in a potential.
In the first case, the motion leads to a characteristic oscillation
frequency related to the tunneling splitting of the ground state energy,
whereas in the second the particle will be find outside{\sc \ }the confining
potential region with a mean scape rate. In this last case, semiclassical
methods are usually employed in order to calculate the mean scape rate and
tunneling probabilities, although other methods are sometimes of great
interest\cite{takagi}.

In what concerns the spin degree of freedom, it has been shown that quantum
tunneling also appears in several physical situations such as magnetic
moments in spin \ glass systems, single magnetic impurity in a crystal
field, a single domain, a ferromagnetic, or antiferromagnetic, domain wall,
and so on\cite{tejada}. In these cases, transitions between states
associated to energetically equivalent configurations exist, even at low
enough temperatures to exclude thermally assisted transitions. Those
transitions were then suggested to be due to quantum tunneling.

From the theoretical point of view, the general approach to the spin
tunneling process was first treated by van Hemmen and S\"{u}to\cite{van},
and by Enz and Schilling\cite{enz}. Further developments came along the
approaches which make use of a generalization of the WKB method adapted to
spin systems\cite{van,walt,zas,van2}, and, alternatively, by using the
Feynman's path integral formulation of quantum mechanics\cite%
{tejada,schilling}. \ \ \ In general, the starting point of any of the two
approaches is a Hamiltonian that describes a spin degree of freedom which
interacts with an external/internal field, for instance,
\[
H=-AS_{z}^{2}+BS_{z}+CS_{x}\text{,}
\]
associated to the manganese acetate complex\cite{tejada}, being these
Hamiltonians written in terms of the standard spin operators obeying the
usual commutation relations

\[
\left[ S_{i},S_{j}\right] =i\varepsilon _{ijk}S_{k}\text{.}
\]

In the present paper we propose an alternative useful approach to the
description of the problem of quantum spin tunneling. The aim of this
approach is to obtain a potential function, written in terms of an angle
variable, which can account for the main properties of the spin system; in
particular we intend to show that this potential can describe the proper
behavior of the spin system when it undergoes a spin tunneling phenomenon.

In order to test our proposal we applied our approach to the
Lipkin-Meshkov-Glick (LMG) model\cite{lipkin}, which is defined through a
many-spins Hamiltonian containing a spin-spin interaction term besides a
one-spin term, i.e., a $J_{z}$ contribution. In fact, that Hamiltonian has
been originally proposed to represent a system of $N$ interacting fermions
in such a form that, due to the algebraic structure involved, it is a
soluble model. It has been also discussed as a {\it bona fide }spin{\it \ }%
model mainly because of its phase transition in the $N\rightarrow \infty $
limit\cite{mar,jaffe}. Here we will be interested in its behavior for finite
values $N$ {\it , }mainly those values which can characterize mesoscopic
systems.

The starting point in the procedure to construct that potential is the
Generator Coordinate Method (GCM) proposed long time ago in order to treat
many-body systems\cite{griwhee} and also adapted to spin systems\cite%
{holz,pipas}. To carry out the calculations we choose the $su(2)$ coherent
states as our generator states and use the LMG Hamiltonian to describe the
spin system. As we get the GCM energy and overlap kernels we construct a
Hilbert space which stands for a new suitable kinematical ground to describe
the spin system collective states. This can be achieved through a procedure
put forth several years ago in order to avoid undesirable diverging
solutions of the original GCM solutions\cite{pipas,pipas2}. Through a
projection of the GCM energy kernel onto this new state space an equation
for the spin system is obtained which presents an energy kernel that is a
nonlocal function of the angular momentum. From this new energy kernel we
can finally extract the potential function in terms of an angle variable\cite%
{garu}.

The behavior of the LMG model gap curve as a function of the two-spin term
interaction strength is then discussed in connection with the associated
potential function, and we show that some aspects of the energy gap curve
can be put in relation to potential curves which exhibit the characteristic
behavior of spin tunneling.

As a further test of the description of this spin system based on a
potential function, we also constructed the Wigner function associated to a
combination of the two lowest states of the model for some particular values
of the strength parameter. For each case its time evolution was then
obtained and the angle distribution probability function extracted. We then
show that the results are also compatible with the potential function
description.

This paper is organized as follows. In section II we recall the LMG model as
a spin system model and discuss some of its properties, mainly those
relevant to the discussion of the spin tunneling. The extraction of the
potential for the spin system is discussed and obtained in section III. In
section IV we present the Wigner function for the two lowest energy states
for different values of the interaction strength parameter and its time
evolution is obtained. Finally section V is devoted to the final
considerations.\qquad

\section{The Lipkin-Meshkov-Glick soluble model}

\subsection{The Lipkin-Meshkov-Glick Hamiltonian}

The Lipkin-Meshkov-Glick (LMG) soluble model\cite{lipkin} was originally
proposed in order to test mean-field approximations in many body systems,
and was introduced as a Hamiltonian containing a one-body term plus a
two-body interaction term in such a way that a set of $N_{p}$ (here
considered an even number) interacting fermions is distributed in two levels
separated by an energy $\epsilon $, being each one of them $N_{p}$-fold
degenerate. The degenerated states within each level are labeled by a
quantum number $r$, which assumes values between $1$ and $N_{p}$, and also
by another quantum number $\sigma $ which is equal to $+1$ or $-1$
designating the upper or the lower level. The Hamiltonian of the model then
reads
\[
H=\frac{\epsilon }{2}\sum_{r,\sigma }\sigma a_{r,\sigma }^{\dagger
}a_{r,\sigma }+\frac{V}{2}\sum_{r,r^{^{\prime }},\sigma }a_{r,\sigma
}^{\dagger }a_{r^{^{\prime }},\sigma }^{\dagger }a_{r^{^{\prime }},-\sigma
}a_{r,-\sigma }.
\]

Although the model was originally tailored for the treatment of fermion
particles, it was soon verified that it could be put in a more simple form
if we realize that a subjacent algebraic structure can be identified within
the operators constituting the Hamiltonian. If we introduce the so-called
quasi-spin operators defined by

\[
J_{+}=\sum_{r}a_{r,+1}^{\dagger }a_{r,-1},
\]

\[
J_{-}=\sum_{r}a_{r,-1}^{\dagger }a_{r,+1},
\]
and

\[
J_{z}=\frac{1}{2}\sum_{r,\sigma }\sigma a_{r,\sigma }^{\dagger }a_{r,\sigma
},
\]
we see that the LMG Hamiltonian can then be written as

\[
H=\epsilon J_{z}+\frac{V}{2}\left( J_{+}^{2}+J_{-}^{2}\right) ,
\]
which exhibits a typical spin character. It is also direct to verify that
those new quasi-spin operators indeed obey the standard $su(2)$ commutation
relations, viz.,
\[
\left[ J_{+},J_{-}\right] =2J_{z},
\]
\[
\left[ J_{z},J_{\pm }\right] =\pm J_{\pm }.
\]

In this way, if we forget for a while its original proposal, the LMG
Hamiltonian can also be studied as a legitimate spin model of its own. The
correspondence between the two operator languages is then direct: In terms
of the fermions it is clear that the $J_{z}$ term gives half the difference
of the number of particles in the upper and lower levels, while the second
term, involving $J_{+}^{2}$ and $J_{-}^{2}$, is associated to the
interaction between a pair of particles in the same energy level, and
scatters this pair to the other level while preserving the quantum number $q$
of each particle. In what follows we will treat the LMG Hamiltonian as a
{\it bona fide }spin{\it \ }model.

It is useful to rewrite that Hamiltonian in such a way to scale the
interaction term to the particle number while measuring the energy in terms
of $\epsilon ,$ i.e.,

\[
H_{L}=\frac{H}{\epsilon }=J_{z}+\frac{\chi }{2N_{p}}\left(
J_{+}^{2}+J_{-}^{2}\right) ,
\]
with $\chi =N_{p}V/\epsilon $.

In what refers to the solutions of this model, we see that from general
symmetry considerations some results can be advanced. First we observe that,
with the Hamiltonian written in terms of these quasi-spin operators, and
since $\left[ H_{L},J^{2}\right] =0$, it can be diagonalized within each
multiplet labeled by the eigenvalues of $J^{2}$ and $J_{z}$, what accounts
for the soluble character of the associated quantum problem. Each multiplet
is $2J+1$-dimensional, and we also note that the ground state belongs to the
multiplet $J=N_{p}/2=\max J_{z}$, being the $J^{2}$ and $J_{z}$ eigenvalues $%
\frac{N_{p}}{2}\left( \frac{N_{p}}{2}+1\right) $ and $-\frac{N_{p}}{2}$,
respectively. Furthermore, we immediately see that the model also presents
two discrete conserved quantities which reflect in the properties of the
solutions. The simplest operator commuting with the Hamiltonian, therefore
giving a constant of motion, is the parity operator

\[
\Pi =\exp \left( i\pi J_{z}\right) ,
\]
indicating that the Hamiltonian matrix, in the $J_{z}$ representation,
breaks into two disjoint blocks involving only even and odd eigenvalues of $%
J_{z}$ respectively. A second interesting property of the energy spectrum is
revealed by the fact that the Hamiltonian anticommutes with the operator

\[
R=\exp \left( i\frac{\pi }{2}J_{z}\right) \exp \left( i\pi J_{y}\right)
=\sum_{m=-N_{p}}^{N_{p}}\mid -m\rangle \left( -1\right) ^{N_{p}+m}\langle
m\mid .
\]
This operator is associated to a rotation of the angular momentum
quantization frame by the Euler angles $\left( -\frac{\pi }{2},\pi ,0\right)
$, thus transforming $H_{L}$ into $-H_{L}$. As a result of this
anticommutation property we observe that if $\mid E_{j}\rangle $ is an
energy eigenstate with eigenvalue $E_{j}$, then the state $R\mid
E_{j}\rangle $ is also an eigenstate of the Hamiltonian with eigenvalue $%
-E_{j}$. This symmetry property of the Hamiltonian gives rise to an energy
spectrum that is symmetric about zero.

\subsection{Some features of the energy gap\ }

In what concerns the exact solutions of the LMG Hamiltonian, they have long
been discussed and found (some of them algebraically) for different $N_{p}$%
's by Lipkin\cite{lipkin}. They can also be numerically obtained by just
diagonalizing the matrix associated to the Hamiltonian in the $J_{z}$ basis.
Once the energy spectrum is obtained it is a simple task to find the energy
gap,$\;\Delta =\left( E_{1}-E_{0}\right) /\epsilon $, which is a very
important quantity characterizing the behavior of this model as a function
of the interaction parameter $\chi $. One of the main features of this model
is that it presents a second order phase transition when $N_{p}\rightarrow
\infty $, as already widely discussed in the literature, as can be seen from
the energy gap curve in the $N_{p}\rightarrow \infty $ limit \ which
presents a typical discontinuity at $\chi =1$, namely
\[
\lim_{N_{p}\rightarrow \infty }\Delta =\left\{
\begin{tabular}{llll}
$\left( 1-\chi ^{2}\right) ^{1/2}$ & $|\chi |$ & $<$ & $1$ \\
$0$ & $|\chi |$ & $>$ & $1$%
\end{tabular}
\right. .
\]
Although this result has a particular interest in the study of systems
presenting phase transitions\cite{jaffe}, here instead, we will be concerned
with the study of the model for finite, medium range values of $N_{p}$ since
it can then stands for a mock mesoscopic spin system.

In this context the main features of the LMG model can be directly extracted
from the energy spectrum. As an illustration of this, Figure 1 shows a
typical energy spectrum for $N_{p}=10$ as a function of the parameter $\chi $%
, the positive part of which is obtained just by reflecting the spectrum
about zero, as mentioned before. As a direct result the energy gap curve can
be obtained and is shown in Figure 2. It is clearly seem that the energy gap
decreases as the parameter $\chi $ increases in such a way that we can infer
the two-body correlation effects introduced by the interaction term of the
Hamiltonian; in other words, as the two-body correlation strength increases
the energy gap decreases. There are, however, as suggested by the shape of
that curve, particular values of the strength parameter $\chi $ which
indicate a change in the effects of the two-body correlations, namely we can
extract the values of $\chi $ for which there occur those changes by just
evaluating the first few derivatives of the energy gap curve. In fact, by
obtaining the first numerical derivatives of the curve shown in Figure 2 we
observe that there are in particular two values of $\chi $, as can be seen
in Figure 3, that indicate those changes. These two values of $\chi $
separate the energy gap curve in three regions -- as also shown in Figure 2.
The first region can be understood as that one in which the mean field --
controlled by the $J_{z}$ term -- dominates, in contrast to the third region
where the two-body correlations are the main responsible for the behavior of
the spectra. The second region is characterized by the competition of both
contributions.

In what follows we will present a description based on a collective
potential extracted from a full quantum mechanical formalism that can be
used to visualize the main aspects of the regions described above.

\section{The LMG collective potential}

As already discussed in Ref. \cite{garu} a procedure can be developed for
the LMG model that permits to treat the quasi-spin system as if it behaves
as a particle system. In other words, we can associate a potential function
of an angle variable defined in the interval $-\pi \leq \phi \leq \pi $ to
describe the spins system interaction.

For the LMG model the whole procedure can be made explicit by firstly
assuming the $su(2)$ coherent states as the generator coordinate method\cite%
{griwhee,pipas} variational space, as has long been proposed\cite{holz},
namely
\[
\mid \alpha \rangle =\cos ^{N_{p}}\left( \frac{\alpha }{2}\right) \exp
\left( \tan \frac{\alpha }{2}J_{+}\right) \mid 0\rangle ,
\]
where $\mid 0\rangle $ is the vacuum of the angular momentum operator. The
generator coordinate method then requires the calculation of an overlap
kernel which, in the present case, is of the form
\begin{equation}
\langle \alpha ^{^{\prime }}|\alpha \rangle =\cos ^{N_{p}}\left( \frac{%
\alpha ^{^{\prime }}-\alpha }{2}\right) .  \label{col2}
\end{equation}
Furthermore, an energy kernel must also be calculated which, in terms of the
new variables $\varphi =\frac{\left( \alpha \acute{}{\acute{}}+\alpha
\right) }{2}$ and $\theta =\alpha \acute{}-\alpha $, reads
\[
\langle \alpha ^{^{\prime }}\left| H_{L}\right| \alpha \rangle =H\left(
\varphi ,\theta \right) =
\]
\begin{equation}
=-\frac{N_{p}}{2}\cos ^{N_{p}}\left( \frac{\theta }{2}\right) \left\{ \cos
\left( \varphi \right) \cos ^{-1}\left( \frac{\theta }{2}\right) +\frac{\chi
}{2}\left[ \cos ^{-2}\left( \frac{\theta }{2}\right) \left( 1+\sin
^{2}\left( \varphi \right) \right) -1\right] \right\} .  \label{col3}
\end{equation}

A new discrete orthonormalized finite-dimensional representation, instead of
this nonorthogonal continuous one, can be achieved\cite{pipas2} if we find a
transformation that diagonalizes the overlap kernel, Equation (\ref{col2}).
This can be directly accomplished by a Fourier transformation
\[
\int_{-\pi }^{\pi }\langle \alpha \acute{}\mid \alpha \rangle \;u_{m}\left(
\alpha \acute{}\right) d\alpha \acute{}=\lambda _{m}u_{m}\left( \alpha
\right) ,
\]
where the eigenfunctions of the overlap kernel are
\[
u_{m}\left( \alpha \right) =\frac{\exp \left( im\alpha \right) }{\sqrt{2\pi }%
}
\]
with the corresponding eigenvalues
\[
\lambda _{m}=\int_{-\pi }^{\pi }\cos ^{N_{p}}\left( \frac{\alpha }{2}\right)
\exp \left( im\alpha \right) d\alpha =\frac{2\pi N_{p}!}{2^{N_{p}}\left(
\frac{N_{p}}{2}+m\right) !\left( \frac{N_{p}}{2}-m\right) !},
\]
with $-\frac{N_{p}}{2}\leq m\leq \frac{N_{p}}{2}.$ Therefore, the
diagonalization process generates a set of $N=N_{p}+1$ orthonormalized
states defining a basis in the ground state multiplet we discussed before. \
The next step consists in projecting the LMG energy kernel, Equation (\ref%
{col3}), onto this angular momentum basis states, which leads to
\[
H\left( m,m\acute{}\right) =\int_{-\pi }^{\pi }\int_{-\pi }^{\pi }\exp \left[
i\varphi \left( m\acute{}-m\right) \right] \exp \left[ -i\frac{\theta }{2}%
\left( m\acute{}+m\right) \right] \frac{H\left( \varphi ,\theta \right) }{%
\sqrt{\lambda _{m}\lambda _{m\acute{}}}}\frac{d\varphi d\theta }{2\pi }.
\]
\ The diagonalization of this matrix gives the exact spectrum of the LMG
Hamiltonian, as expected.

Since the angular momentum space is characterized by a finite set of states,
we can obtain a new discrete representation of the LMG Hamiltonian, viz.,
the {\it discrete angle representation} associated to the angular momentum
one through a discrete Fourier transform, namely,
\[
H(k,k\acute{})=\sum_{m=-\frac{N-1}{2}}^{\frac{N-1}{2}}\sum_{m\acute{}=-\frac{%
N-1}{2}}^{\frac{N-1}{2}}\frac{\exp \left( im\theta _{k}\right) }{\sqrt{N}}%
H\left( m,m\acute{}\right) \frac{\exp \left( -im\acute{}\theta _{k\acute{}%
}\right) }{\sqrt{N}},
\]
where now $\theta _{k}$ labels the discrete angle variable, and $\theta _{k}=%
\frac{2\pi k}{N}$.

The final entangled matrix of the energy kernel in this discrete angle
representation does not allows us a direct comparison with Equation (\ref%
{col3}), written in terms of the continuous angle. However, the
finite-dimensional character of this description can be exploited to allow
for a discrete phase space description of this kind of degree of freedom\cite%
{gapi1} in which an angle-angular momentum pair labels each point of the
space. One virtue of this phase space picture is that it further allows for
the extraction of a potential function, described in terms of the angle
variable, much in the same spirit as for particle systems, which can then
describe the behavior of the spin system. This extraction can be implemented
if we realize that the zeroth moment of this matrix, with respect to the
variable $\theta _{k}-\theta _{k\acute{}}$, represents a potential function
in the discrete angular variable $\left( \theta _{k}+\theta _{k\acute{}{%
\acute{}}}\right) /2$. To this end, let us define
\[
\theta _{k}-\theta _{k\acute{}}=u,
\]
and
\[
\theta _{k}+\theta _{k\acute{}}=2\phi .
\]
Due to the periodicity of the functions involved, we must only consider the
interval for the summation over $k$ to run as $-\frac{N-1}{2}\leq k\leq
\frac{N-1}{2}$. Furthermore, in order to emphasize the angular character of
this discrete variable we will consider that range to be defined as $-\frac{%
N-1}{N}\pi \leq u\leq \frac{N-1}{N}\pi $, being the steps of this variable $%
\Delta u=\frac{2\pi }{N}$.

In this form, the zeroth moment will be calculated as
\[
M_{0}\left( \phi \right) =\sum_{u=\text{\/}-\frac{N-1}{N}\pi }^{\frac{N-1}{N}%
\pi }\frac{\Delta u}{2\pi }\sum_{m,m\acute{}=-\frac{N_{p}}{2}}^{\frac{N_{p}}{%
2}}\int_{-\pi }^{\pi }\int_{-\pi }^{\pi }\frac{\exp \left[ i\left( \phi
-\varphi \right) \left( m-m\acute{}\right) \right] }{2\pi \sqrt{\lambda
_{m}\lambda _{m\acute{}}}}
\]
\begin{equation}
\times \exp \left[ i\left( u-\theta \right) \frac{\left( m\acute{}+m\right)
}{2}\right] H\left( \varphi ,\theta \right) d\varphi d\theta .  \label{col4}
\end{equation}
Defining the new variables $r=m\acute{}-m$ and $s=\frac{\left( m\acute{}%
+m\right) }{2}$, and using the general result
\[
\sum_{m=-\frac{N_{p}}{2}}^{\frac{N_{p}}{2}}\sum_{m\acute{}=-\frac{N_{p}}{2}%
}^{\frac{N_{p}}{2}}=\sum_{r=-N_{p}}^{-1}\sum_{s=-N_{p}-r}^{N_{p}+r}+%
\sum_{r=1}^{N_{p}}\sum_{s=r-N_{p}}^{N_{p}-r}+\sum_{r=0,s=-N_{p}}^{N_{p}},
\]
where the summations over $s$ are restricted to run only over even(odd)
values depending if $r$ is even(odd), we can perform all the integrals and
summations. Therefore, Equation (\ref{col4}) gives us the discrete potential
function associated to the Lipkin model in terms of the angular variable $%
\phi $.

It has been also shown \cite{garu} that the large $N_{p}$ limit can be
obtained from this approach which coincides with the result by Holzwarth\cite%
{holz}. The form for the potential, when $N_{p}>>1$, is given by
\[
V\left( \phi \right) =M_{0}\left( \phi \right) =-\frac{\left( N_{p}-1\right)
}{2}\cos \left( \phi \right) -\frac{\chi \left( N_{p}+3\right) }{4}\sin
^{2}\left( \phi \right) .
\]
It is interesting to realize that the Lipkin model attains a classical limit
when $N_{p}\rightarrow \infty $, in the sense that quantum dynamics becomes
the classical dynamics\cite{jaffe}. Although this result guides us in the
study of the behavior of large systems, for small or intermediate values of $%
N_{p}$, when the two-body correlations are as important as the mean field
term, and must be taken into account, the potential function must be
calculated directly from Equation (\ref{col4}) which keeps all information
regarding the spin system.

Now we can look at the potential functions associated to the LMG model
calculated for those special values of the interaction parameter $\chi $,
obtained by the first three derivatives of the gap curve, as mentioned
before. In Figure 4 we show the various curves associated to the potential
for different values of $\chi $ for $N_{p}=10$; it is also shown in each
case the ground state and the first excited state energies, which give the
energy gap. It is immediately seen that for $\chi =0$, when the spectrum
comes from a pure $J_{z}$ term, the potential presents a single minimum at $%
\phi =0$. The bottom of the potential then becomes more and more shallow as $%
\chi $ increases leading to a narrowing of the energy gap, i.e., the energy
levels approach each other due to the uncertainty principle since the
potential is then more wide in the angle variable. This behavior
characterizes the first region in the gap curve. Now, the potential changes
from a minimum to a maximum at $\phi =0$ when $\chi \simeq 0.85$ for $%
N_{p}=10$. Soon after the maximum at the origin appears, more specifically
around $\chi \simeq 1.2$, the first particular value of the interaction, a
more pronounced barrier starts to show up at the origin, and the wave
functions present reflection effects so that the energy gap changes its
behavior; for this value of $\chi $ starts the second region. We then
observe that for the second value of $\chi $, namely $\chi \simeq 1.8$, the
second energy level touches the top of the barrier, and tunneling begins to
play an essential role in defining the energy gap. Therefore, the second
region ends while the third region starts which consists essentially of spin
tunneling through a barrier. In Figure 4 we also show the plot for $\chi
=2.5 $, a paradigmatic case of the third region, where tunneling fully takes
place.

Thus, we see that a potential function written in terms of an angle variable
can be constructed for the LMG quasi-spin model in such a way that its
description of the system is consistent with the exact results, obtained
from the direct diagonalization of the energy matrix in the angular momentum
representation, and can then account for a natural description of the
behavior of the energy levels as the interaction strength changes.
Furthermore, this potential function description may then stands for a
natural tool to treat spin tunneling processes.

\section{The time evolution of the Wigner function}

Another approach to the LMG model energy gap curve comes from the study of
the time evolution of the Wigner function associated to the lowest energy
levels. Since the finite-dimensional phase spaces basic ideas have already
been presented in the past\cite{gal1,gapi1}, we briefly recall how this
scheme can be implemented.

First of all we identify the finite set of states of the angular momentum
which labels the LMG model ground state multiplet $|\frac{N_{p}}{2},m\rangle
$ as the eigenstates of the Schwinger\cite{schw} unitary operator $U$, i.e.,
\begin{equation}
U\mid \frac{N_{p}}{2},m\rangle =\exp \left( \frac{2\pi i}{N}m\right) \mid
\frac{N_{p}}{2},m\rangle ,  \label{wig1}
\end{equation}
in such a way that we can assume $\mid \frac{N_{p}}{2},m\rangle \equiv
\;\mid u_{m}\rangle $.

Since we know the LMG quasi-spin Hamiltonian, the mapped Liouville operator
which governs the time evolution of the Wigner function on the discrete
phase space -- labelled by the angular momentum and angle pair $(m,n)$ --
can be directly calculated by means of the mapping
\[
{\cal L}\left( u,v,r,s\right) =\frac{1}{N}Tr\{G^{\dagger }(u,v)\left[ H,-%
\right] \},
\]
where
\[
G\left( m,n\right) =\sum_{j,l}\frac{U^{j}V^{l}}{N}\exp \left[ i\pi \beta
\left( j,l;N\right) -\frac{2\pi i}{N}\left( mj+nl\right) +i\frac{\pi }{N}jl%
\right] ,
\]
is the operator basis proposed in\cite{gapi1}, $U$ and $V$ are the Schwinger
unitary cyclic shifting operators\cite{schw}, and the phase $\beta \left(
j,l;N\right) $ performs all the mod $N$ arithmetic involved in the given
operator basis calculations.

It has been shown\cite{gal1} that the discrete mapped Liouvillian reads
\[
{\cal L}\left( u,v,r,s\right) =2i\sum_{m,n}\sum_{a,b,c,d}\frac{h\left(
m,n\right) }{N^{4}}\sin \left[ \frac{\pi }{N}\left( bc-ad\right) \right]
\exp \left[ i\pi \Phi \left( a,b,c,d;N\right) \right]
\]
\[
\exp \left\{ \frac{2\pi i}{N}\left[ a\left( u-m\right) +b\left( v-n\right)
+c\left( u-r\right) +d\left( r-s\right) \right] \right\} ,
\]
where $h\left( m,n\right) $ is the discrete phase space mapped form of the
LMG quasi-spin Hamiltonian, namely
\[
h\left( m,n\right) =m+\frac{\chi }{N_{p}}\sqrt{\left( \frac{N_{p}}{2}%
+m\right) \left( \frac{N_{p}}{2}+m+1\right) \left( \frac{N_{p}}{2}-m\right)
\left( \frac{N_{p}}{2}-m+1\right) }\cos \frac{2\pi }{N}2n,
\]
and $\Phi \left( a,b,c,d;N\right) =-\beta \left( a+c+\frac{N_{p}}{2},b+d+%
\frac{N_{p}}{2};N\right) $.

We also recall that the von Neumann-Liouville time evolution equation for
the density operator, for time independent Hamiltonians,
\[
i\frac{\partial }{\partial t}\widehat{\rho }\left( t\right) =\left[ H,%
\widehat{\rho }\left( t\right) \right] ,
\]
is represented in the discrete phase space by the mapped expression
\begin{equation}
i\frac{\partial }{\partial t}\rho _{w}\left( u,v;t\right) =\sum_{r,s}{\cal L}%
\left( u,v,r,s\right) \rho _{w}\left( r,s;t\right) ,  \label{wig6}
\end{equation}
where now $\rho _{w}\left( u,v;t\right) $ is the Wigner function associated
to the state of the system we are interested in\cite{gapi1}. A solution to
Equation (\ref{wig6}) in the form of a series
\[
\widehat{\rho }\left( t\right) =\widehat{\rho }\left( t_{0}\right) +\left(
-i\right) \left( t-t_{0}\right) \left[ H,\widehat{\rho }\left( t_{0}\right) %
\right] +\frac{1}{2!}\left( -i\right) ^{2}\left( t-t_{0}\right) ^{2}\left[ H,%
\left[ H,\widehat{\rho }\left( t_{0}\right) \right] \right] +...
\]
has its discrete phase space mapped expression written as\cite{garu}
\[
\rho _{w}\left( u,v;t\right) =\sum_{r,s}\left\{ \delta _{r,u}^{\left[ N%
\right] }\delta _{s,v}^{\left[ N\right] }+\left( -i\right) \left(
t-t_{0}\right) {\cal L}\left( u,v,r,s\right) +\right.
\]
\begin{equation}
+\left( -i\right) ^{2}\left( t-t_{0}\right) ^{2}\sum_{x,y}\left. \frac{1}{2!}%
{\cal L}\left( u,v,x,y\right) {\cal L}\left( x,y,r,s\right) +...\right\}
\rho _{w}\left( r,s;t_{0}\right) .  \label{wig8}
\end{equation}

Therefore, since we are given the Liouvillian for the LMG model, we can
obtain the Wigner function at any instant of time $t$ by using the series
representing the iterated application of the discrete mapped Liouville
operator. It is also{\sc \ }to be noted that this time evolution is then
carried out as a linear process of composition of sums of products of the
arrays characterizing the Liouvillian and the Wigner functions respectively,
over all the sites constituting the associated discrete phase space of the
model. From an operational point of view, we have to construct the arrays
defining the Liouvillian for the LMG model and the Wigner function. Once the
Wigner function is given at time $t_{0}$, the calculations with the series,
Equation (\ref{wig8}), will give us the propagated Wigner function at any
instant of time.

Furthermore, as has been already shown, the probability distribution for the
angular momentum and for the angle variables, as a function of time, can be
obtained directly from the Wigner function by means of a trace operation,
\[
L\left( m;t\right) =\sum_{n=-\frac{N_{p}}{2}}^{\frac{N_{p}}{2}}\rho
_{w}(m,n;t),
\]
and
\begin{equation}
\Phi \left( n;t\right) =\sum_{m=-\frac{N_{p}}{2}}^{\frac{N_{p}}{2}}\rho
_{w}(m,n;t)  \label{wig10}
\end{equation}
respectively. It must be noted that the angle variable is also discrete
being characterized by $\theta _{n}=\frac{2\pi }{N}n$.

\subsection{Wigner functions associated to some of the LMG quasi-spin model
states}

Since the general mapping expression
\[
O\left( m,n\right) =\frac{1}{N}Tr\left[ G^{\dagger }\left( m,n\right) O%
\right]
\]
holds for any operator $O$ acting on the states of the corresponding state
space, we will be concerned with the density operator $\rho $, in particular
for pure states
\[
\rho =\;\mid \psi \rangle \langle \psi \mid ,
\]
when these states can be expressed in terms of the basis states of the
unitary $U$ operator, Equation (\ref{wig1}), which are also eigenstates of
the $J_{z}$ operator, namely,
\[
\mid \psi \rangle =\sum_{k}c_{k}\mid u_{k}\rangle .
\]
For these cases it is simple to show\cite{gapi1} that the Wigner function
reads
\begin{equation}
\rho _{w}\left( m,n\right) =\frac{1}{N^{2}}\sum_{j,l=-\frac{N_{p}}{2}}^{%
\frac{N_{p}}{2}}\exp \left[ \frac{2\pi i}{N}\left( mj+nl\right) \right]
f\left( j,l\right) ,  \label{wig12}
\end{equation}
where
\[
f\left( j,l\right) =\sum_{p=-\frac{N_{p}}{2}}^{\frac{N_{p}}{2}%
}c_{[p+l]}^{\ast }c_{p}\exp \left[ -\frac{2\pi i}{N}j\left( p+\frac{l}{2}%
\right) \right] ,
\]
and the index $[p+l]$ just stands for the value of the sum $p+l$ that are
cyclically restricted to the range of the labels, e.g., $p_{\max }+1=p_{\min
}$. Furthermore, since we have chosen the ranges of the labels $j,l$ and $p$
to be symmetric about the origin, we will obtain real-valued Wigner
functions, and due to this choice it must be also noted that, in this
context, the $%
\mathop{\rm mod}%
\;N$ phase $\beta (m,n;N)$ present in the operator basis will not contribute
to the above expression$.$ The advantage of having Wigner functions in that
form is that if we diagonalize the LMG quasi-spin Hamiltonian in the $J_{z}$
basis we can obtain directly the coefficients $c_{k}$ in order to build a
Wigner function using Equation (\ref{wig12}). In particular, we can
construct the symmetric and antisymmetric combinations
\[
\mid \psi ^{s}\rangle =\frac{1}{\sqrt{2}}\sum_{k=-\frac{N_{p}}{2}}^{\frac{%
N_{p}}{2}}\left[ c_{k}^{i}\mid u_{k}^{i}\rangle +c_{k}^{j}\mid
u_{k}^{j}\rangle \right] ,
\]
and
\[
\mid \psi ^{a}\rangle =\frac{1}{\sqrt{2}}\sum_{k=-\frac{N_{p}}{2}}^{\frac{%
N_{p}}{2}}\left[ c_{k}^{i}\mid u_{k}^{i}\rangle -c_{k}^{j}\mid
u_{k}^{j}\rangle \right]
\]
from the i-th and j-th eigenstates of the LMG quasi-spin model respectively.
These states will be of great interest specially when dealing with the
ground and the first excited state since then we know that the energy gap
curve as a function of $\chi $ indicates the existence of three regions of
interest for finite $N_{p}$. From those two states we immediately obtain, by
means of the direct use of Equation (\ref{wig12}), the two corresponding
Wigner functions, i.e., $\rho _{w}^{s}\left( m,n\right) $ and $\rho
_{w}^{a}\left( m,n\right) $ respectively.

To carry out the time evolution of the Wigner function for any one of these
possibilities, we have only to apply Equation (\ref{wig8}). Furthermore, it
is also direct to see that we can extract the superposition probability of
the $t=0$ state, represented by the corresponding Wigner function
(symmetric/antisymmetric), with a time evolved one. This will indicate us
the existence of periods of oscillation between symmetric/antisymmetric
states for specified values of the interaction parameter $\chi $. To obtain
this probability as a function of time we have to realize that
\begin{equation}
P_{if}\left( t\right) =\sum_{m,n}\rho _{w}^{i}\left( m,n\right) \rho
_{w}^{f}\left( m,n;t\right) ,  \label{III8}
\end{equation}
so that we have only to carry out the summations over the whole discrete
phase space of the product of the Wigner functions.

All the numerical calculations were carried out for $1200$ steps of time,
each one being $\Delta t=0.1$, and the family of curves corresponding to the
symmetric case is presented in Figure 5. In all of them we started with the
symmetric Wigner function at $t=0$, and calculated the probability of
obtaining the symmetric Wigner function again at any time $t$. It is clear
that for $\chi =0.0$, when the Hamiltonian is a pure $J_{z}$ term, and the
symmetric state only involves the ground state and its closest neighbor with
no coupling, the corresponding Wigner function will oscillate back and forth
with period $2\pi $, since the energy gap, $\Delta $, is exactly $1$ (it
must be recalled that we are measuring energies in units of $\epsilon $).
This corresponds simply to a free oscillation, since the spectrum is equally
spaced, there are no coupling term in the Hamiltonian, and no tunneling at
all. We then observe that, as soon as the interaction strength $\chi $
begins to increase, the oscillations change their frequencies in such a form
that the basic relation
\[
\omega (\chi )=\frac{\Delta \left( \chi \right) }{2}
\]
is verified, as is expected for the particular case of a mixture of two
energy eigenstates of the Hamiltonian.

As a by-product of the time evolution of the Wigner function, we can also
obtain the mean value of the phase variable, $\phi $, of the system state at
any instant of time by just considering the average of the phase
distribution function, Equation (\ref{wig10}), with the corresponding Wigner
function, in the same form as that prescribed in Equation (\ref{III8}). The
results for the symmetric case are presented in Figure 6. From those plots
it can be clearly seen that while the Wigner function oscillates between the
symmetric and antisymmetric configurations periodically in time, the mean
phase bounces back and forth in the potential. As $\chi $ increases and the
barrier at the origin begins to play a role in the time evolution, the
oscillation frequency in the potential diminishes and the Wigner function
spends more and more time in one of the two minima, until it tunnels to the
other one. For sufficiently high values of $\chi $ the frequency tends to
zero (since $\Delta $ goes to zero), and the angle distribution keeps
trapped into one of the two minima of the potential therefore stopping the
tunneling process.

\section{Conclusions}

In this paper we have shown that a potential function description for the
quasi-spin Lipkin-Meshkov-Glick model can be constructed that can account
for all properties a typical spin system presents. Although originally based
in the treatment of many particle systems, this description has been
partially developed and exploited in the past, and shown here to be of use
for spin systems as well. In fact, we have explored the LMG model for finite
intermediate number of spins in order to simulate mesoscopic spin systems.
As such, we have shown how to describe some interesting features of the
energy gap curve exhibited in that model for $N=10$ in terms of the behavior
of the spin system phase state in a potential. In particular this
description allows us to interpret the appearance of three distinct regions
in the energy gap curve as the change of behavior of that phase state when
the form of the potential changes as the interaction parameter $\chi $
increases. When $\chi $ exceeds a certain value, we are led to a typical
tunneling process in the proposed potential that can then be fully described
in terms of a phase variable. In this sense the proposed potential is an
alternative useful tool to handle spin tunneling processes.

We also have shown that a discrete Wigner function associated to a symmetric
(or antisymmetric) combination of the ground and first excited states of the
LMG model can be constructed and its time evolution obtained when
considering the proposed potential. The time behavior of the Wigner function
corroborates our description of the LMG quasi-spin model. The phase mean
value can also be directly obtained from the Wigner function and clearly
shows the oscillations of the phase state in the potential. When $\chi $
greatly increases, leading to the presence of two pronounced minima, the
frequency of oscillations becomes lower and lower until the phase state is
completely trapped in one of the two minima.

\acknowledgements{\it The authors would like to thank Prof. W. F. Wreszinski
for important suggestions, and to draw our attention to the paper by R.
Schilling. B.M.P. and D.G. acknowledge partial financial support from
Conselho Nacional de Desenvolvimento Cient\'{i}fico e Tecnol\'{o}gico, CNPq,
Brazil. B.M.P. also acknowledges Funda\c{c}\~{a}o de Amparo \`{a} Pesquisa
do Estado de S\~{a}o Paulo, FAPESP, Brazil, Proc. n}$^{o}${\it \
2002/00222-9. C.L.L. acknowledges Funda\c{c}\~{a}o de Amparo \`{a} Pesquisa
do Estado de S\~{a}o Paulo, FAPESP, Brazil, Proc. n}$^{o}${\it \
2002/10896-7.}

\newpage

\bigskip {\bf Figure Captions}

{\bf Figure 1}: Energy spectrum for the Lipkin model with N =10 particles as
a function of the strength parameter $\chi $. The positive part of the
spectrum is obtained by reflecting the negative part around zero.

\bigskip {\bf Figure 2}: The gap function $\Delta =E_{1}-E_{0}$ as a
function of the strength parameter $\chi $.

\bigskip {\bf Figure 3}: First three numerical derivatives of the gap curve
presented in the previous figure. The vertical bars indicate the relevant
values of $\chi $ where the correlations change the gap in a peculiar way.
In this figure the solid line indicates the first derivative ($n=1$), the
dashed line corresponds to $n=2$, and the long dashed-short dashed line
curve to $n=3$.

\bigskip {\bf Figure 4}: Family of potential curves associated to
different values of the strength parameter $\chi $. The curve for $\chi
=0$ shows the potential for a spin system without two-body
correlations. The curves for $\chi =1.2$ and $\chi =1.8$ show the
potential for the particular values indicated in the previous figure.
The curve for $\chi =2.5$ exhibits a typical case of spin tunneling.

\bigskip {\bf Figure 5}: Family of curves showing the probability of finding
the symmetric combination as a function of time for some values of the
strength parameter $\chi $. We have prepared a symmetric combination at
$t=0$. As $\chi $ increases the frequency diminishes.

\bigskip {\bf Figure 6}: Family of curves showing the mean phase value as a
function of time for some values of the strength parameter $\chi $ for the
symmetric combination. It is shown that, as $\chi $ increases, the system
state bounces with a lower frequency between the two wells.

\begin{figure*}[h]
{\centerline{\epsfbox{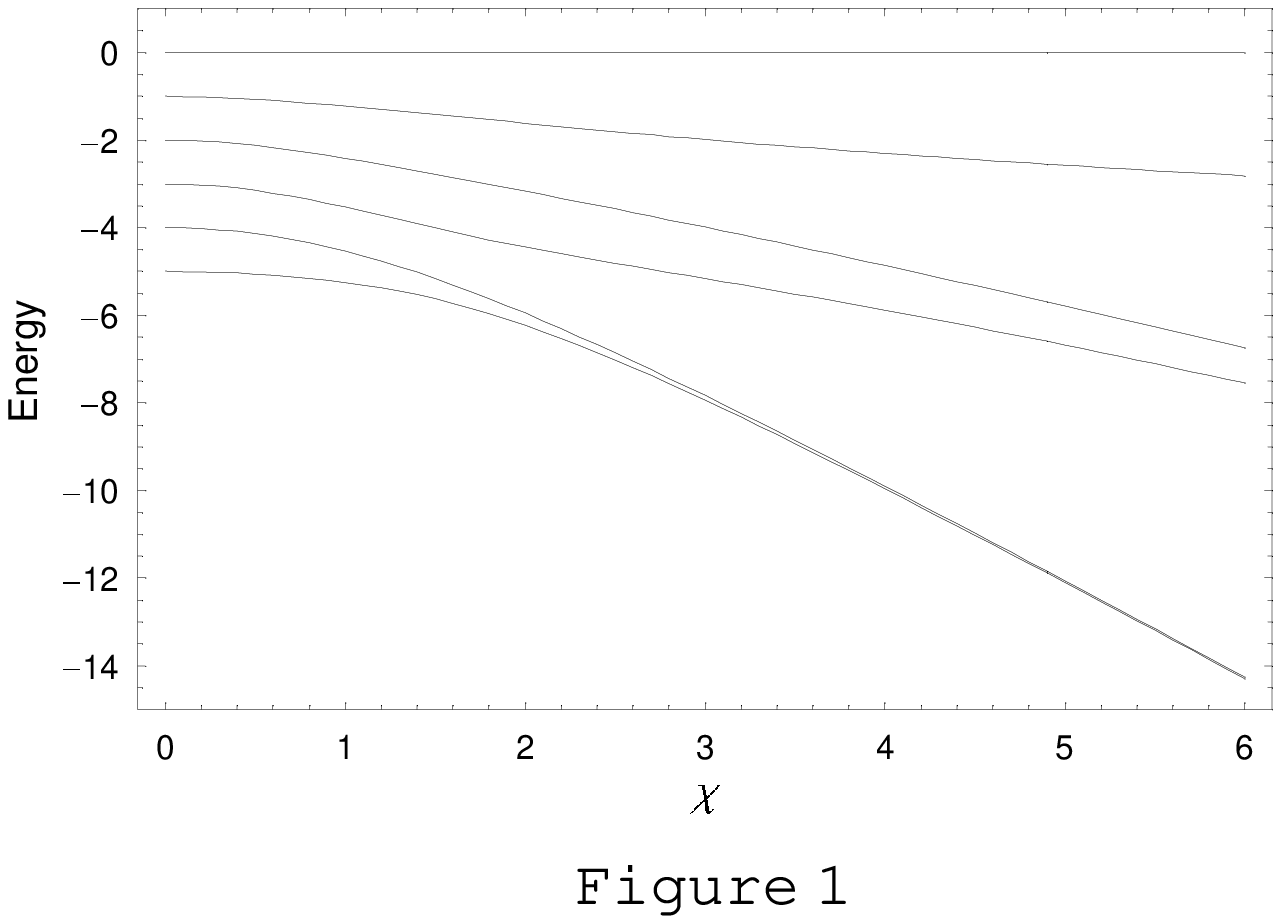}}}
\end{figure*}

\begin{figure}[h]
{\centerline{\epsfbox{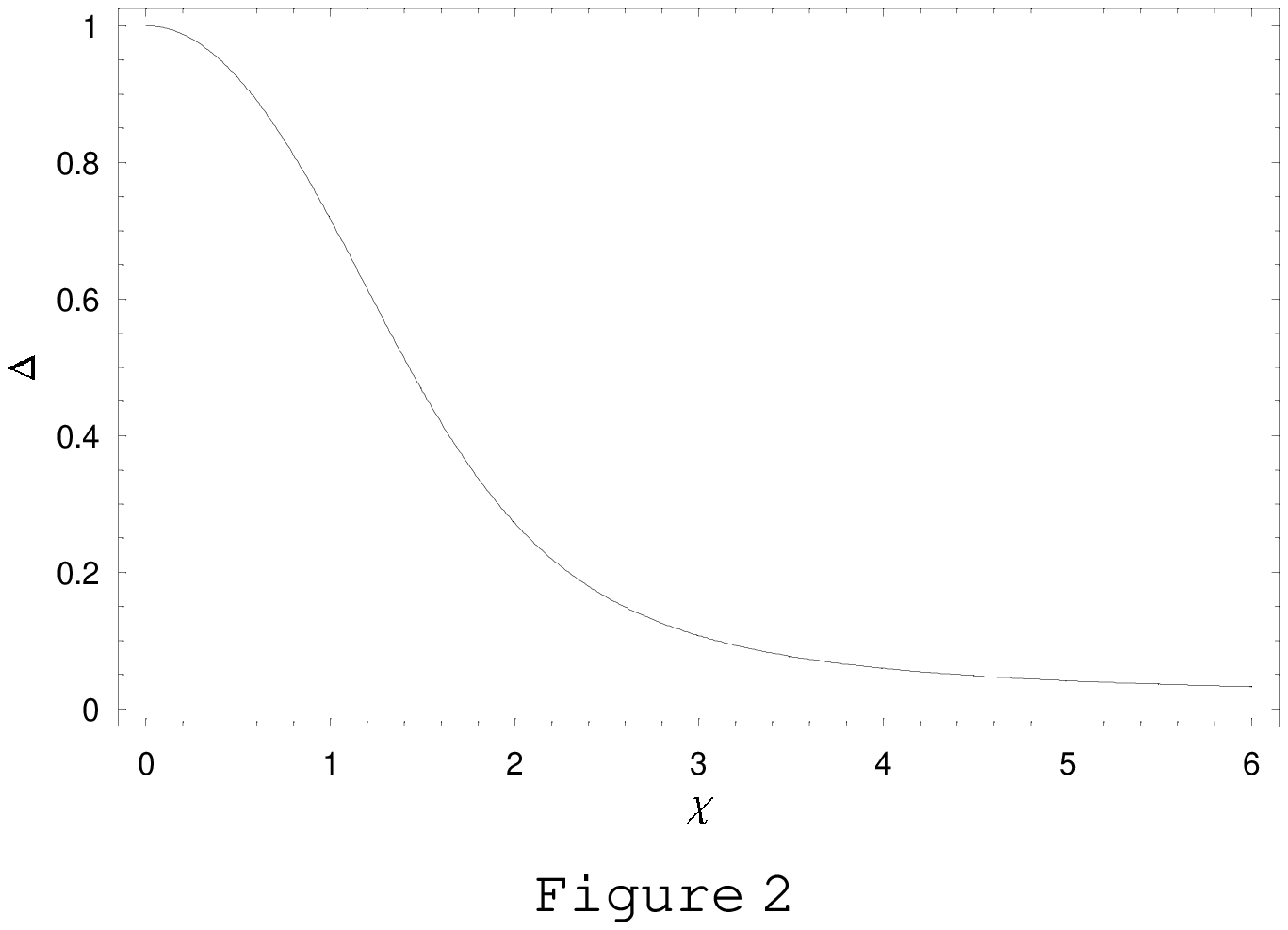}}}
\end{figure}
\begin{figure}[h]
{\centerline{\epsfbox{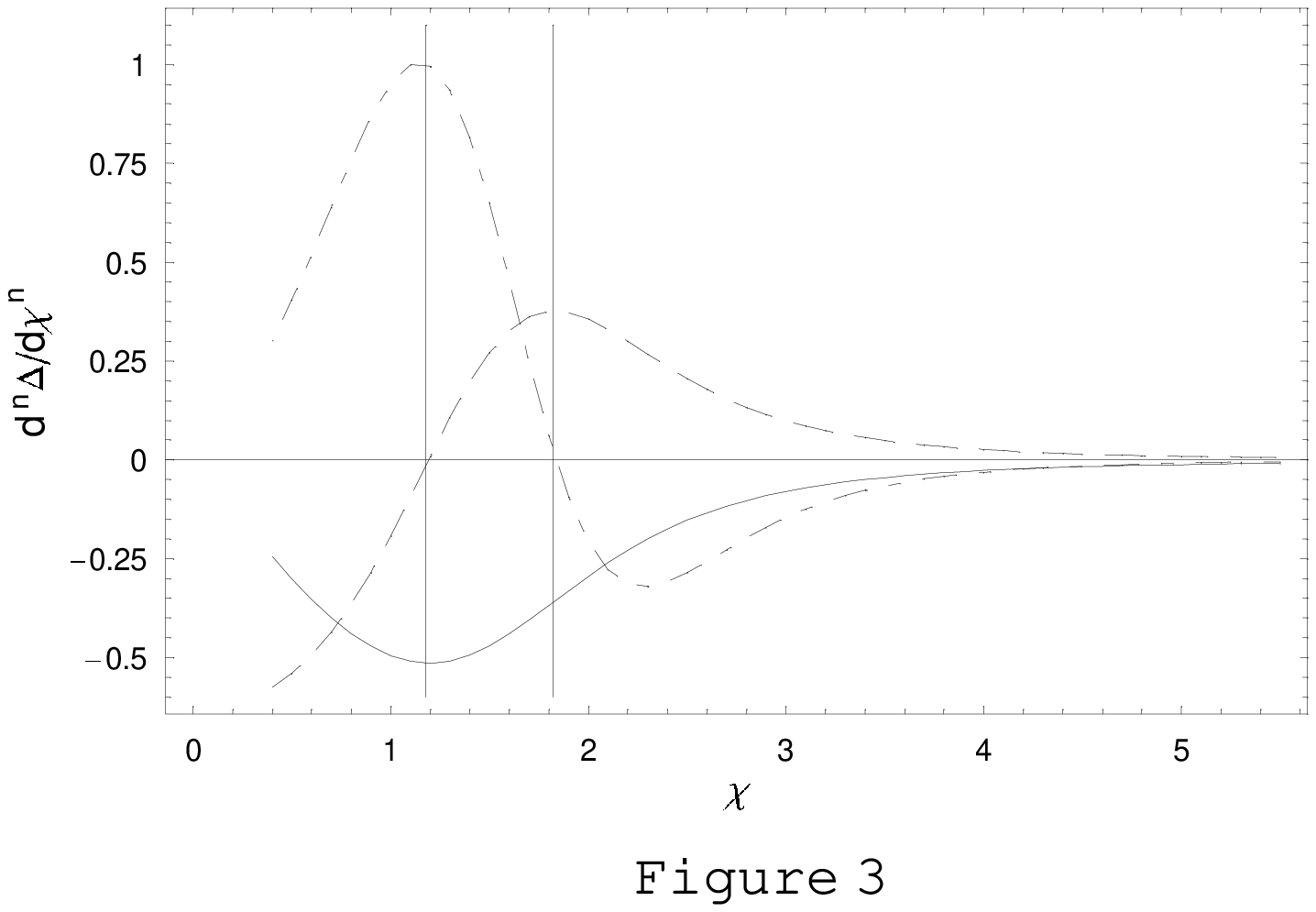}}}
\end{figure}
\begin{figure}[h]
{\centerline{\epsfbox{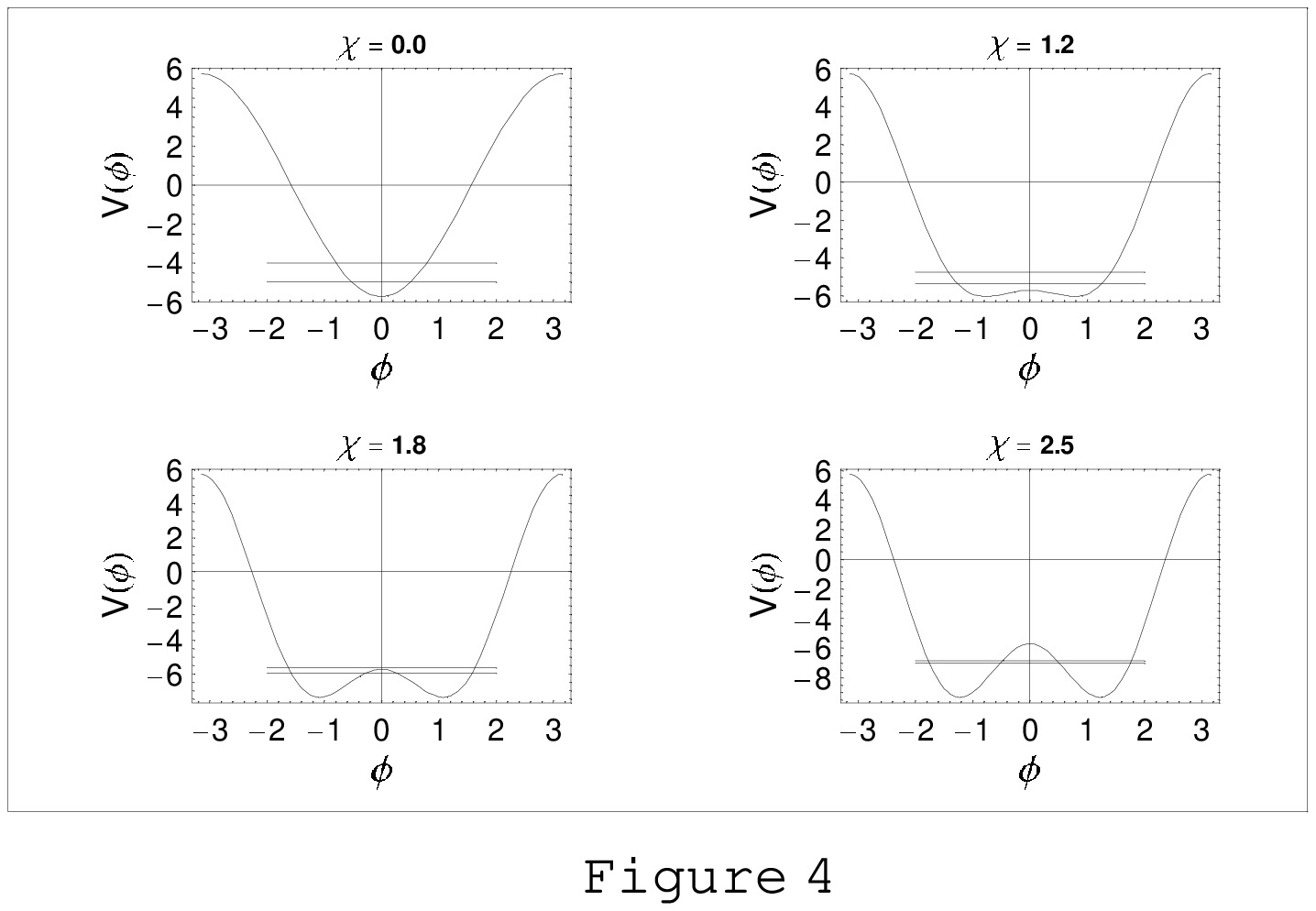}}}
\end{figure}
\begin{figure}[h]
{\centerline{\epsfbox{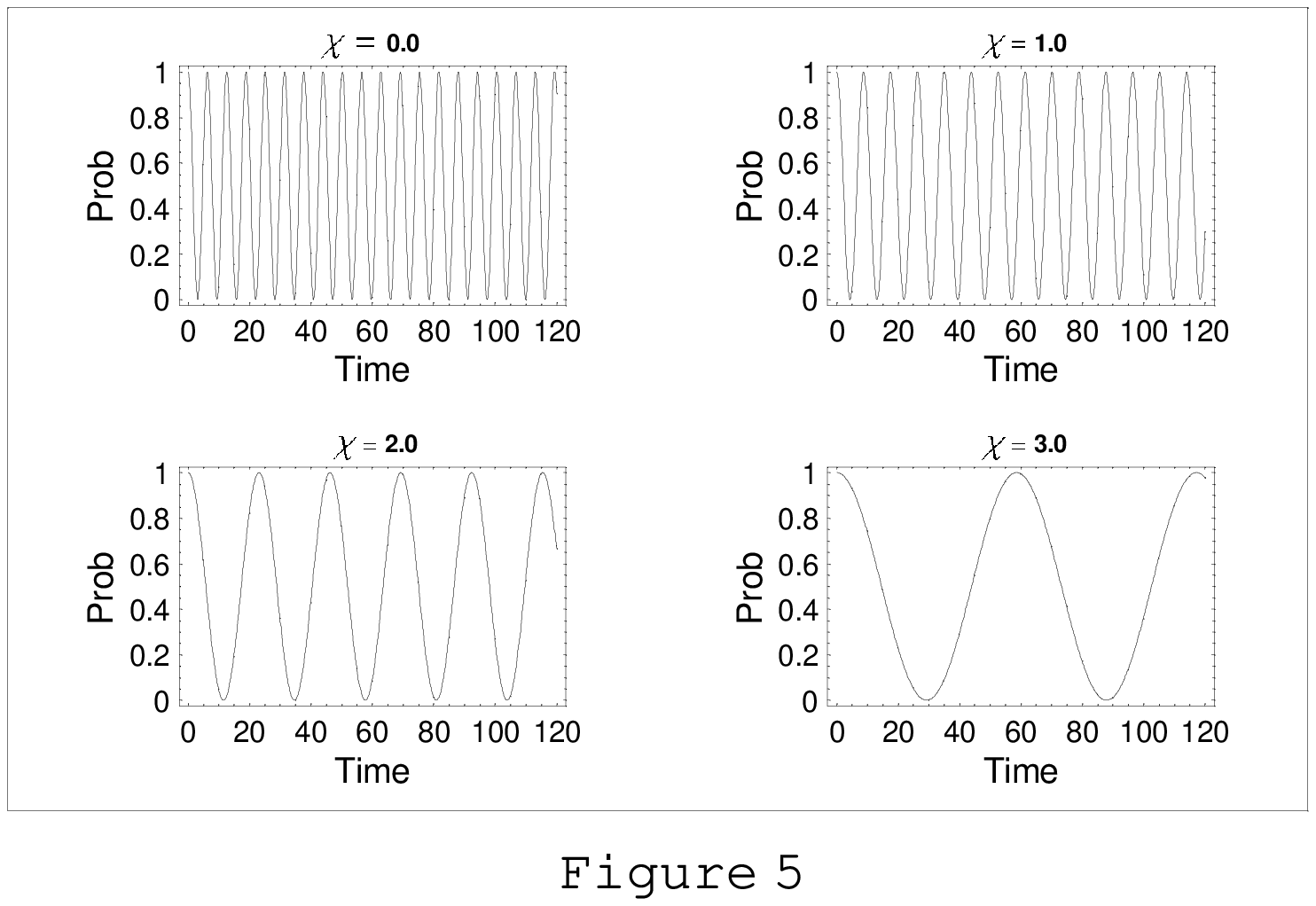}}}
\end{figure}
\begin{figure}[h]
{\centerline{\epsfbox{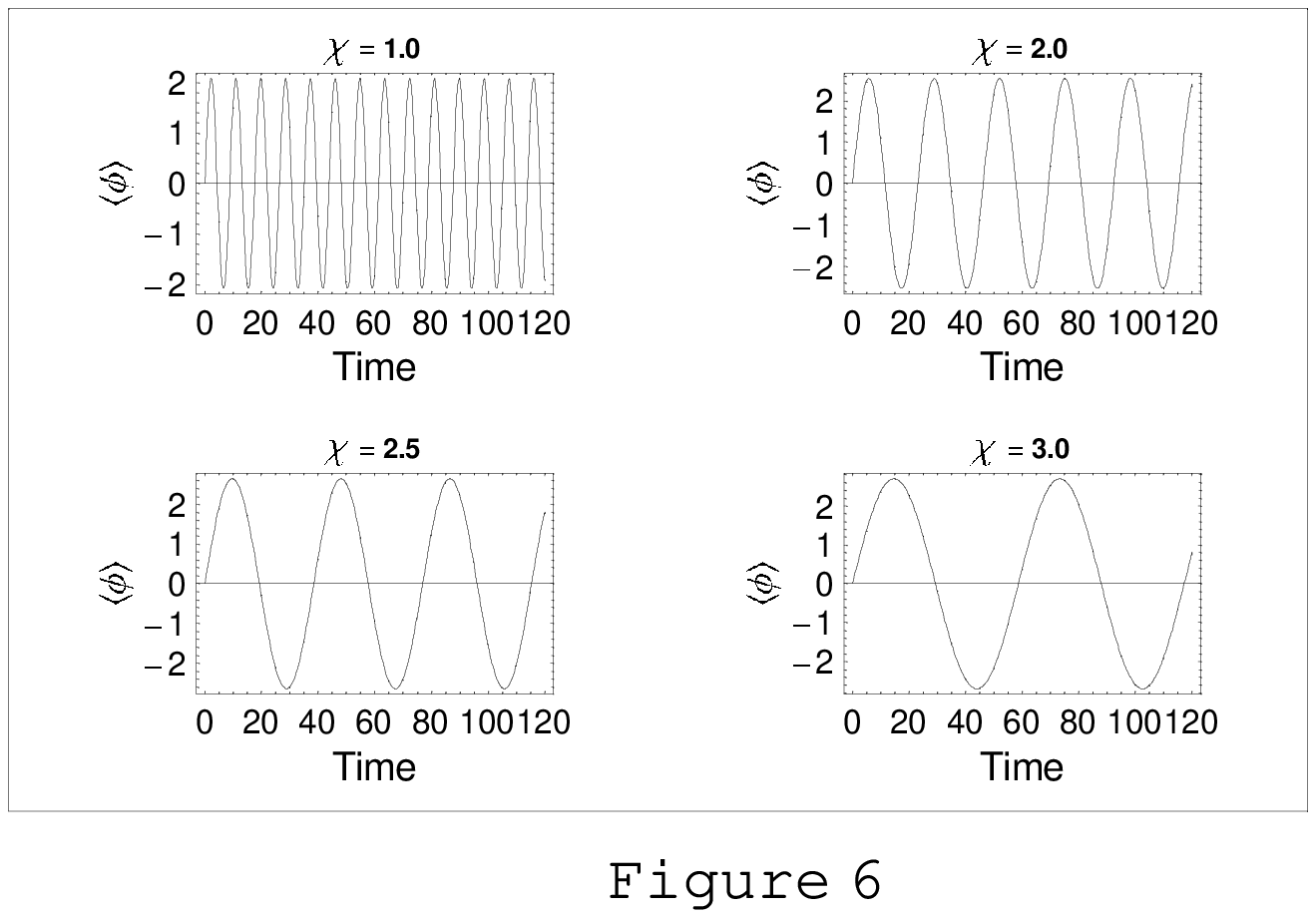}}}
\end{figure}
\end{document}